\documentstyle[12pt]{article}

\input epsf

\newcommand{\beq}{\begin{equation}}
\newcommand{\eeq}{\end{equation}}
\newcommand{\bea}{\begin{eqnarray}}
\newcommand{\eea}{\end{eqnarray}}

\begin{document}
{\bf\hfil\break
 DOE-ER40757-094\hfil\break
UTEXAS-HEP-97-3}\hfil\break \hfil\break
\begin{center}
{\Large {\bf Finite Temperature Wave-Function Renormalization, 
A Comparative Analysis}}
\end{center}
\hfil\break
\begin{center}
{\bf Ian A. Chapman}
\end{center}
\begin{abstract}
We compare two competing theories regarding finite temperature wave-function 
corrections for the process $H\rightarrow e^+e^-$ and for $n+\nu\rightarrow
p+e^-$ and related processes of interest for primordial nucleosynthesis.  
Although the two 
methods are distinct (as shown in $H\rightarrow e^+e^-$) they yield 
the same finite temperature correction for all $n\rightarrow
p$ and $p\rightarrow n$ processes.  Both methods yield an increase in the 
He/H ratio of $.01\%$ due to finite temperature renormalization rather
than a decrease of $.16\%$ as previously predicted.

\end{abstract}

\section{Introduction}

To calculate finite temperature corrections to physical processes
beyond the tree 
level, one needs to have a consistent and correct method of 
temperature dependent renormalization.  Unfortunately, there seems to be more
than one way to do this. 
Although everyone seems to agree on how to handle temperature dependent 
phase space, temperature dependent vertex corrections, effective mass 
corrections, and temperature dependent bremsstrahlung corrections, they do not 
agree on how one handles wave function renormalization.

In this paper, we will examine two competing models.  Each 
model's effect on the 'effective' projection operator (which in turn will tell
us how to handle wave function renormalization) is examined for the general
case.  These general results are then applied to $H\rightarrow e^+e^-$ 
since this is a 
fairly simple process to understand.  We also apply both methods to 
processes of interest for primordial nucleosynthesis, specifically 
$n+\nu\rightarrow p+e^-$ and related processes.  These processes were first 
analyzed by Dicus,et al.\cite{D}, and independently by Cambier, Primack, 
and Sher\cite{C}.  
Since these papers were published, however, it has been claimed that neither 
paper handled the finite temperature wave function renormalization scheme 
correctly.  Both Donoghue and Holstein\cite{DH}, and Sawyer\cite{S} propose 
models that claim 
to correct this difficulty.  Both agree that one must use {\sl finite 
temperature} Dirac spinors (and thus projection operators), but otherwise 
their models seem to differ significantly.

Before we begin a few comments about gauges are in order.  The correction of 
interest will be the wavefunction renormalization part of the width.  
Since this is a gauge dependent quantity, we might wonder if our choice of 
gauge will make a difference in our results.  We solve this problem in two 
ways.  For most of this paper, we will avoid a specific choice of gauge.  
Instead we will discuss the differences between the models in the most general
 way possible.  When we calculate a specific result for nucleosynthesis, 
this is no longer a viable option.  In Appendix I, however, we prove that 
we can talk about the differences in the wavefunction renormalization parts 
in the width in a gauge invariant manner.  We take advantage of this in 
Eq.(14).   

\section{Generalized Analysis}

As discussed in the previous literature, the presence of a background thermal
bath breaks Lorentz invariance.  At T=0 we can identify the wavefunction 
renormalization constant Z$_2$ for fermions with momentum $p$ as
 1-Z$_2=\left .{\partial\Sigma\over{\partial\not p}}\right |_{{\not p}=m}$
where $\Sigma$ is the self energy. 
Since Lorentz invariance is broken when T$\neq$0, it is claimed that 
this identification of a 
momentum independent renormalization constant is no longer correct.  
Donoghue and Holstein advise us to look at the modified propagator 
instead 
and to modify the Dirac spinors accordingly.  Thus the modified propagator
becomes
\beq G={1\over\not p-m-\Sigma} \eeq
where the fermion self-energy, $\Sigma$ is
\beq\Sigma=Ap_0\gamma^0-Bp_i\gamma^i-C.\eeq
$A$,$B$, and $C$ are momentum dependent scalar functions.  We 
identify an "effective" momentum, ${\tilde p}^\mu$ where ${\tilde p}^\mu=
((1-A(p))p^0,(1-B(p))p^i)$ and an effective mass $\tilde m$ where 
${\tilde m}= m-C(p)$.  We are instructed to replace the fermion momenta 
and mass with the "effective" fermion momenta and mass everywhere inside the 
Dirac spinors.  When this is done, {\sl and} the projection operator is 
expanded about the {\it physical} mass shell energy, we get the following 
"effective" projection operator
\beq\Lambda^+_{eff}={\not\tilde p+\tilde m\over{2\tilde p_0}}\eeq 

Converting this into the mass shell $p^\mu$, expanding about the mass-shell
energy, and writing this in terms of $A$,$B$, and $C$, we get (noting that
$\omega_p=\sqrt{{\bf p}^2+m^2}$ or the mass shell energy)

\beq\Lambda^+_{eff}=\left(1-\left .{{\partial 
((1-2A)p_0^2-(1-2B){\bf p}^2-(1-{2C\over m})
m^2)\over\partial p_0}\over 2\omega_p (1-A)}\right |_{p_0=\omega_p}\right)
\left({\not p+m\over 2\omega_p}\right)\eeq 

Simplifying this, we find

\beq\Lambda^+_{eff}=\left(A+\left .
{\partial A\over\partial p_0}\right |_
{p_0=\omega_p}\omega_p-\left .{\partial B\over\partial p_0}\right|_{
p_0=\omega_p}{{\bf p}^2\over\omega_p}-\left .
{\partial C\over\partial p_0}\right|_{
p_0=\omega_p}{m\over\omega_p}\right)\left({\not p+m\over 2\omega_p}\right)
\eeq

Thus the effective projection operator is equal to the original 
T=0 projection operator multiplied by a single scalar momentum dependent 
function.  Donoghue and Holstein identify this function as 
1-"Z$_2$" where "Z$_2$" is the effective temperature dependent 
wave-renormalization factor.  

Sawyer gives us a differing way to handle finite temperature wave-function 
renormalization. Sawyer tells us to start with the same modified propagator as
shown in Eq.(1), but then reminds us that the propagator can also be 
written in terms of the effective Dirac spinors as shown by Weldon\cite{W}.  
Thus \beq G={\Sigma_{\alpha}w_{\alpha}{\bar w}^{\alpha}\over p_0-
\omega_p}+
finite\;terms\eeq
where $w$ is the temperature dependent renormalized Dirac spinor.  Using this 
fact, Sawyer is able to perturbatively derive the temperature dependent 
renormalized Dirac 
spinors, $w$, in terms of the unrenormalized zero temperature Dirac spinors, 
$u$.  He was then able to use these new spinors to directly 
define the effective projection operators.  If we follow this procedure we 
get
\beq\Lambda^+_{eff}={1\over 4\omega_p}
(\gamma^0\left.{\partial\Sigma\over\partial p_0}
\right|_{p_0=\omega_p}(\not p+m)+(\not p+m)\left.{\partial\Sigma\over\partial
 p_0}
\right|_{p_0=\omega_p}\gamma^0)\eeq
When we write Sawyer's effective projection operator in terms of $A$,$B$,$C$,
and $\omega_p$ as defined previously, we get (evaluating $p_0$ at $\omega_p$).
$$\Lambda^+_{eff}=\left(A+{\partial A\over\partial p_0}
p_0\right)\left(
{\not p+m\over 2\omega_p}\right)\mskip200mu$$
\beq +\left({\partial B\over\partial p_0}p_i
\gamma^ip_0-{\partial B\over\partial p_0}{\bf p}^2\gamma^0-{\partial C\over
\partial p_0}p_0-{\partial C\over\partial p_0}m\gamma^0\right)\left({1\over
2\omega_p}\right)\eeq 
It is immediately evident that the above expression for $\Lambda^+_{eff}$ is 
in general different from the expression for $\Lambda^+_{eff}$ in Eq.(5).
Specifically, we note that $\Lambda^+_{eff}$ is {\it not} equal to the 
unrenormalized projection operator times some scalar functional factor.  
We see this explicitly in the process $H\rightarrow e^+e^-$.

\section{$\bf H\rightarrow e^+e^-\;comparison$}

The tree-level matrix element for this process in momentum space is
\beq M_T=-{igm\over2M_W}{\bar u}(e^-)v(e^+)\eeq
where $m$ is the mass of the fermion, $M_W$ is the mass of the $W^{\pm}$, 
${\bar u}$ and $v$ are the Dirac Spinors at T=0, and $M_T$ is the tree level
matrix element for this process. 
The modified width considering {\sl only} those terms due to the 
temperature dependent wavefunction renormalization can be found (to first order
in $\alpha$) by cross-multiplying the tree-level matrix elements using
the modified projection operators, $\Lambda^+_{eff}$ and $\Lambda^-_{eff}$, in 
place of the normal zero temperature unrenormalized projection operators.  One 
then integrates over the correct temperature dependent phase space.  
Since this later part is not in dispute, it is easier to simply compare the 
modified spin-summed squared self-energy matrix elements in both models.  (As 
I mentioned before, I will neglect the temperature dependent mass terms here.)
We assume for this process that the Higgs is at rest with respect to the
background thermal bath, and that the Higgs is significantly more massive than
the background temperature.  When we use the modified projection operator for 
Donoghue and Holstein's model, we get
\beq\left|M\right|^2={2M^2_Hg^2m^2(1-{4m^2\over M^2_H})
\over4M^2_W}
\left(A+{\partial A\over\partial e_0}e_0-{\partial B\over\partial e_0}{{\bf e}
^2\over e_0}-{\partial C\over\partial e_0}{2m\over M_H}\right)\eeq
When we use the modified projection operator as described by Sawyer,
 however, we get
\beq\left|M\right|^2={g^2m^2\over4M^2_W}
\left[2M^2_H\left(1-{4m^2\over M^2
_H}\right)\left(A+{\partial A\over\partial e_0}e_0\right)+{\partial C\over
\partial e_0}mM_H\right]\eeq
These matrix elements squared yield differing widths when integrated 
over phase space.  Thus the two models are distinct for $H\rightarrow e^+e^-$.

\section{\bf $n+\nu\rightarrow p+e^-$ comparison}

Finite temperature corrections have been of considerable interest for $n
\rightarrow p$ and $p\rightarrow n$ reactions because of nucleosynthesis.  
Specifically, one would like to know  
how finite temperature corrections affect abundance ratios such as H/He in
 the early universe.  Since all $n\rightarrow p$ and $p\rightarrow n$ processes
 are related by virtually identical matrix elements, a comparison of 
$n+\nu\rightarrow p+e^-$ is sufficient for all such processes.

In this comparison we make the following standard assumptions (valid 
for proton to neutron ratios in the early universe).  First we assume that
the ambient temperatures of interest are of order  1 MeV  
so that the proton and neutron have considerably more mass than the 
ambient temperature.  (Thus any terms of the form $m\over M_p$ or 
$m\over M_n$ are ignored.)  We also assume that the initial nucleon is at 
rest with 
respect to the surrounding medium.  The neutrino electron part of the 
spin-summed squared matrix element  
 looks like
\beq {\rm Tr}\left\{\Lambda^+_{eff}{1\over2}\gamma^{\mu}
(1-\gamma^5){\not\!\nu}{1\over2}
(1+\gamma^5)\gamma^{\nu}\right\}\eeq

In general $\Lambda^+_{eff}$ has three kinds of terms.  There are scalar terms,
terms that are multiplied by a factor of $\gamma^0$, and the remaining terms
which are all multiplied by a factor of $e_i\gamma^i$.  The scalar terms 
will trace away, and those terms that are multiplied by $e_i\gamma^i$ do not 
survive the phase space integration.  Thus the {\it only} terms in the 
effective projection operators that matter are those terms which are multiplied
by $\gamma^0$. We now look back at Donoghue and Holstein's effective 
projection operator (Eq.(5)) and Sawyer's effective projection operator 
(Eq.(8)).  Isolating the $\gamma^0$ terms in both models, we get
\beq \Lambda^+_{eff}=\left(A+{\partial A\over\partial e_0}e_0
-{\partial B\over
\partial e_0} {e_v^2\over e_0}-{\partial C\over\partial e_0}{m\over e_0}
\right){e_0\gamma^0}\eeq
i.e., the effective projection operators are {\it identical}. 
In the above equation, $e_v$ is the magnitude of the 3-momentum of the 
electron.   To verify this, I did this 
calculation using Donoghue and Holstein's method 
{\sl and} Sawyer's method 
using the Feynman gauge.   (Sawyer used the Coulomb gauge in his 
analysis.) 

When we calculate the width according to Sawyer (and Donahue and 
Holstein) minus
 the width according to Dicus et al. (and Cambier et al.) 
for the reaction
$n+\nu\rightarrow p+e^-$ we get
$$\delta\Gamma_{Sawyer}-\delta\Gamma_{Dicus}={G_F^2(g_V^2+3g_A^2)\alpha
\over 2 \pi^4}\int_Q(Q-e)^2e_v{\rm d}e\int_0k^2{\rm d}k$$
$$\left[1+exp(-e/T)\right]^{-1}\left[1+exp({e-Q\over T})\right]^{-1}
\times \left\{{n_f(E_k)\over E_k}\left[{4m^2(e+E_k)\over 
(E_ke+m^2)^2-e_v^2k^2}\right.\right.$$ $$\left.\left.+{4m^2(e-E_k)\over 
(E_ke-m^2)^2-e_v^2k^2}+{E_k\over{e_v}k}{\rm Ln}\left|{(E_ke-
e_vk)^2-m^4\over (E_ke+e_vk)^2-m^4}\right|\right.\right.$$ 
\beq\left.\left.+{e\over e_vk}{\rm Ln}
\left|{(E_k^2e^2-(e_vk-m^2)^2\over E_k^2e^2-(e_vk+m^2)^2}\right|
\right]-2{n_b(k){\rm Ln}\left|{e+e_v\over e-e_v}\right|\over e_vk}
\right\}
\eeq
where $Q$ is the rest mass of the neutron minus the rest mass of the proton, 
$e_v=\sqrt{e^2-m^2}$, $E_k=\sqrt{k^2+m^2}$, $n_f(E_k)=[1+exp(E_k/T)]^{-1}$
, $n_b(k)=[exp(k/T)-1]^{-1}$.
   Using the above equation and its analogs 
for the other five $n\rightarrow p$ and $p\rightarrow n$ reactions, we obtain
 the change in the width ratios in figures one and two.  

In figure 
three we combine the $n\rightarrow p$ and $p\rightarrow n$ processes into 
a single ratio change for $n\rightarrow p$ and $p\rightarrow n$.  

We use 
these results to solve for the change in the asymptotic neutron to baryon 
ratio,
 $\delta X_n$ as described by Bernstein\cite{B} where
\beq\delta X_n(t)=S(t)\int^t{\rm d}t'\left[\delta\Gamma (t')_{p\rightarrow n}
[1-X_n(t')]-\delta\Gamma (t')_{n\rightarrow p}X_n(t')\right]S^{-1}(t')
\eeq
where $S(t)$ is the solution of ${{\rm d}S\over{\rm d}t}+(\Gamma_{n\rightarrow p}+
\Gamma_{p\rightarrow n})S=0$.   

We solve Eq.(15) for the $\delta X_n$ change we expect.  We solve 
for $S(t)$, $\delta\Gamma_{p\rightarrow n}$, and $\delta\Gamma_{n\rightarrow
 p}$ using Vegas (a Monte-Carlo numeric integration routine).  We solve the
 outer integration over $t$ by Gaussian quadratures.  We obtain an 
increase of $\delta X_n$ of 0.000415.  Thus in Dicus et al.'s original 
paper, $Y_3-Y_4$ in Table 1 (on page 2701) should be changed from +0.0004 to 
-0.000015.  
This corresponds with an increase of $.01\%$ in $\delta X_n$ due to finite 
temperature renormalization as compared to a $.16\%$ decrease as predicted 
by Dicus et al. and Cambier et al.  Sawyer predicts an 
increase of $.02\%$ in $\delta X_n$ due to finite temperature renormalization.
  Although his result differs somewhat from ours, we both agree that the 
change in $\delta X_n$ should be very small and positive i.e., increase due  
to temperature dependent renormalization effects.

\section{Summary}

It is clear that the models proposed by Donoghue and Holstein, 
and Sawyer are 
indeed {\sl distinct}.  However, it is equally clear that both models 
yield the same finite temperature corrections for 
$n\rightarrow p$ and $p\rightarrow n$ processes when the temperature is much  
less than the nucleon mass.  This is because the later 
processes, at least as described in the early universe, are insensitive to any
 correction that does not depend directly on the $\gamma^0$ or energy component
 of the effective projection operator.  It is extremely interesting to note 
that the effective projection operators for both models are {\it identical} 
in this one crucial component.  

If one can show that in one's process of interest, the $\gamma^i$ and scalar 
pieces of the renormalized projection operator don't matter, 
then you can 
indeed use Donoghue and Holstein's effective "Z$_2$" in the 
conventional way.  Otherwise, it is not clear which method, if either,
should be used.  

\section{Acknowledgments}

I especially wish to thank D.A. Dicus for his assistance.  Without his 
patience, insight, support, and encouragement, this research would not have 
been possible.  I also thank M. Byrd and X. Bonnin for discussions and 
technical assistance.  This research was supported in part by the U.S. 
Department of Energy under Contract \# DE-FG3-93ER40757.

\hfil\break
\hfil\break
\hfil\break

\section{Appendix I}

We want a gauge independent way of comparing two finite temperature 
renormalization models.  It is not clear that looking solely at the 
wavefunction or self-energy part of the renormalized width will be sufficient
 since this part is gauge dependent.  Fortunately we know that all of the 
overall models we consider in this paper are gauge invariant {\it and} 
that the only part of these models that differ are the self-energy parts.  
Thus
\beq \delta\Gamma^{total}_{Sawyer}=\delta\Gamma^{SE}_{Sawyer}
+\delta\Gamma^{other}_{Sawyer}\eeq
and
\beq \delta\Gamma^{total}_{Dicus}=\delta\Gamma^{SE}_{Dicus}
+\delta\Gamma^{other}_{Dicus}\eeq
In the above equations, $\delta\Gamma^{total}$ represents the total change
 in width due to finite temperature renormalization, $\delta\Gamma^{SE}$
 represents the change in width due to finite temperature wavefunction 
renormalization, while $\delta\Gamma^{other}$ represents all other changes 
in the width due to finite temperature renormalization.  Since the models 
proposed by Dicus and Sawyer {\sl only} disagree on the wavefunction 
renormalization part, we conclude that $\delta\Gamma^{other}$ must be the 
same for both models.  Thus if we know subtract Eq.(17) from Eq.(16) we get
\beq \delta\Gamma^{total}_{Sawyer}-\delta\Gamma^{total}_{Dicus}=
\delta\Gamma^{SE}_{Sawyer}-\delta\Gamma^{SE}_{Dicus}\eeq
Since we {\it know} the total change in widths is gauge independent for 
both models, it follows from Eq.(18) that $\delta\Gamma^{SE}_{Sawyer}-
\delta\Gamma^{SE}_{Dicus}$ must be gauge independent as well.  This gives us
 a good way to compare two different models for finite temperature 
renormalization.

\begin{figure}
\centerline{\epsfxsize=5in \epsfbox{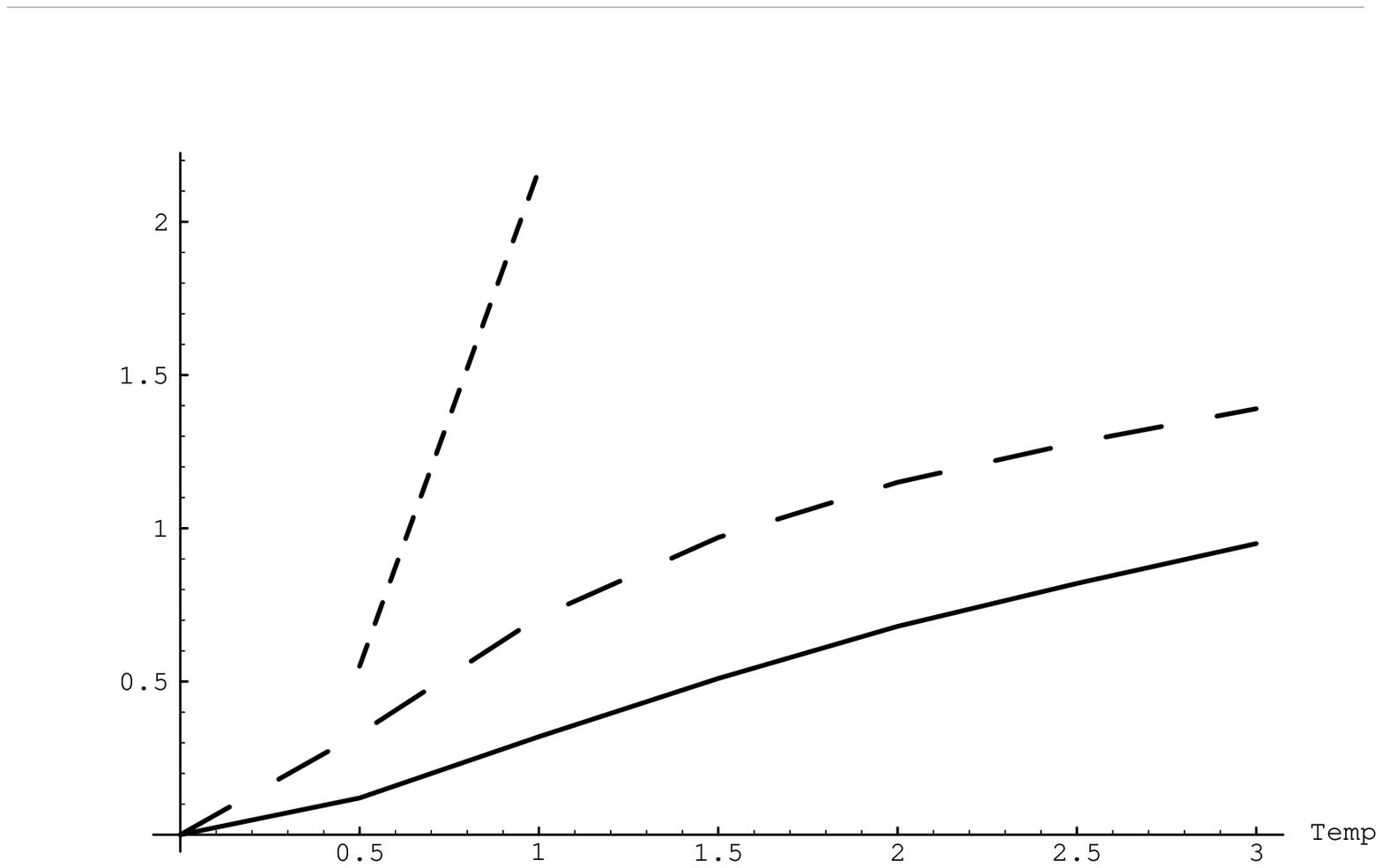}}
\caption{These are the ratios of the change of the finite temperature first 
order corrections (as calculated in Eq.(14)) to the finite temperature 
uncorrected width for all $n\rightarrow p$ processes in units of 
${\alpha\over\pi}$. The temperature 
is measured in units of the electron mass.  The short dashed line 
represents $n\rightarrow pe\nu$, the long dashed line represents $ne
\rightarrow p\nu$, and the solid line represents $n\nu\rightarrow pe$.}
\end{figure}
\begin{figure}
\centerline{\epsfxsize=5in \epsfbox{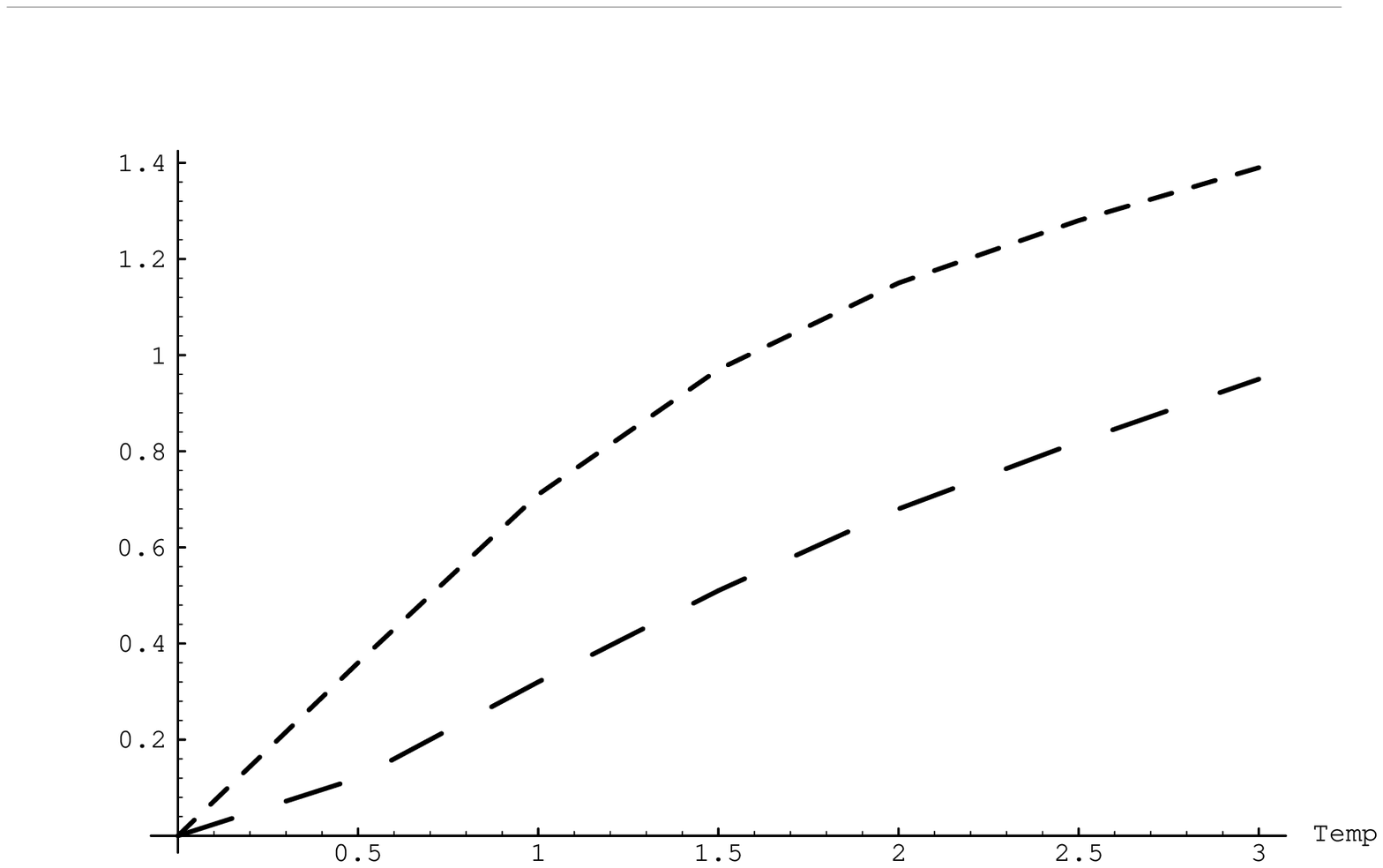}}
\caption{These are the ratios of the change of the finite temperature first 
order corrections (as calculated in Eq.(14)) to the finite temperature 
uncorrected widths for all $p\rightarrow n$ processes in units of 
${\alpha\over\pi}$. The temperature 
is measured in units of the electron mass.  The short dashed line 
represents $p\nu\rightarrow ne$, and the long dashed line represents 
$pe\rightarrow n\nu$}
\end{figure}

\begin{figure}
\centerline{\epsfxsize=5in \epsfbox{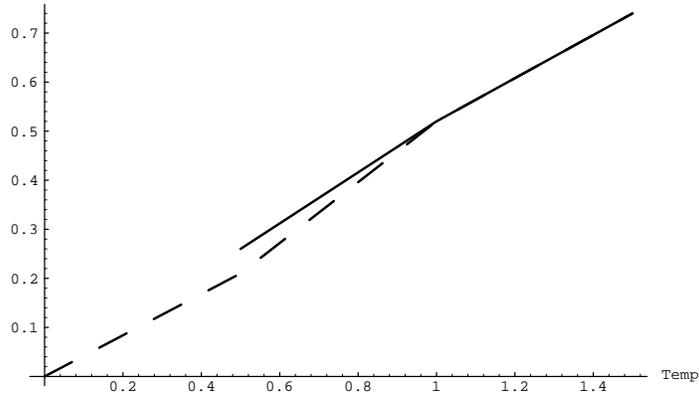}}
\caption{These are the combined changes in the finite temperature first order
 corrections (as calculated in Eq.(14)) to the combined finite temperature
 uncorrected widths for both $n\rightarrow p$ and $p\rightarrow n$ in units 
of ${\alpha\over\pi}$. The 
temperature is measured in units of the electron mass.  The solid line 
represents the combined $n\rightarrow p$ processes while the dashed line 
represents the combined $p\rightarrow n$ processes.}
\end{figure}

\end{document}